\documentclass[aps,showpacs,preprintnumbers,twocolumn]{revtex4}
\usepackage{amsmath,amssymb,graphicx}

\def\be{\begin{equation}}
\def\ee{\end{equation}}
\def\bea{\begin{eqnarray}}
\def\eea{\end{eqnarray}}
\newcommand{\nn}{\nonumber}

\def \l {\lambda}

\def \d {\delta}

\def \g {\gamma}

\def \e {\epsilon}

\def \5 {{}^{(5)}}

\begin{document}

%BA-TH 619-09

\title{Classical tests of general relativity in brane world models}

\author{Christian G. B\"ohmer}
\email{c.boehmer@ucl.ac.uk}\affiliation {Department of Mathematics
and Institute of Origins, University College London, Gower Street,
London, WC1E 6BT, UK}

\author{Giuseppe De Risi}
\email{giuseppe.derisi@ba.infn.it}\affiliation{Dipartimento di
Fisica, Universit\'a degli studi di Bari} \affiliation{Istituto
Nazionale di Fisica Nucleare, sez. di Bari,\\ Via G. Amendola 173,
70126, Bari, Italy}

\author{Tiberiu Harko}
\email{harko@hkucc.hku.hk} \affiliation{Department of Physics and
Center for Theoretical and Computational Physics, The University
of Hong Kong, Pok Fu Lam Road, Hong Kong}

\author{Francisco S. N. Lobo}
\email{flobo@cii.fc.ul.pt} \affiliation{Centro de Astronomia e
Astrof\'{\i}sica da Universidade de Lisboa, Campo Grande, Ed. C8
1749-016 Lisboa, Portugal}

\date{\today}

\begin{abstract}

The classical tests of general relativity (perihelion precession,
deflection of light, and the radar echo delay) are considered for
several spherically symmetric static vacuum solutions in
brane world models. Generally, the spherically symmetric vacuum
solutions of the brane gravitational field equations, have
properties quite distinct as compared to the standard black hole
solutions of general relativity. As a first step a general
formalism that facilitates the analysis of general
relativistic Solar System tests for any given spherically
symmetric metric is developed. It is shown that the existing
observational Solar System data on the perihelion shift of
Mercury, on the light bending around the Sun (obtained using
long-baseline radio interferometry), and ranging to Mars using the
Viking lander, constrain the numerical values of the parameters of
the specific models.

\end{abstract}


\maketitle

\section{Introduction}

The idea, proposed in \cite{RS99a}, that our four-dimensional
Universe might be a three-brane, embedded in a five-dimensional
space-time (the bulk), has attracted considerable interest in the
past few years. According to the brane world scenario, the
physical fields (electromagnetic, Yang-Mills, etc) in our
four-dimensional Universe are confined to the three-brane. These
fields are assumed to arise as fluctuations of branes in string
theories. Only gravity can freely propagate in both the brane and
bulk space-times, with the gravitational self-couplings not
significantly modified. This model originated from the study of a
single $3$-brane embedded in five dimensions, with the 5D metric
given by $ ds^{2}=e^{-f(y)}\eta _{\mu \nu }dx^{\mu }dx^{\nu
}+dy^{2}$, which due to the appearance of the warp factor $f(y)$, could
produce a large hierarchy between the scale of particle physics
and gravity. Even if the fifth dimension is uncompactified,
standard 4D gravity is reproduced on the brane. Hence this model
allows the presence of large, or even infinite non-compact extra
dimensions. Our brane is identified to a domain wall in a
5-dimensional anti-de Sitter space-time. Due to the correction
terms coming from the extra dimensions, significant deviations
from the Einstein theory occur in brane world models at very high
energies \cite{SMS00,SSM00}. Gravity is largely modified at the
electro-weak scale $1$ TeV. The cosmological and astrophysical
implications of the brane world theories have been extensively
investigated in the literature \cite{all}. Gravitational collapse
can also produce high energies, with the five dimensional effects
playing an important role in the formation of black holes
\cite{all1}.

In standard general relativity the unique exterior space-time
of a spherically symmetric object is described by the Schwarzschild
metric. In the five dimensional brane world models, the high energy
corrections to the energy density, together with the Weyl stresses
from bulk gravitons, imply that on the brane the exterior metric
of a static star is no longer the Schwarzschild metric
\cite{Da00}. The presence of the Weyl stresses also means that the
matching conditions do not have a unique solution on the brane;
the knowledge of the five-dimensional Weyl tensor is needed as a
minimum condition for uniqueness.
Static, spherically symmetric exterior vacuum solutions of the
brane world models were first proposed in \cite{Da00} and in
\cite{GeMa01}. The first of these solutions, obtained in
\cite{Da00}, has the mathematical form of the Reissner-Nordstrom
solution of standard general relativity, in which a tidal Weyl
parameter plays the role of the electric charge of the general
relativistic solution. The solution was obtained by imposing the
null energy condition on the 3-brane for a bulk having non zero
Weyl curvature, and it can be matched to the interior solution
corresponding to a constant density brane world star. A second
exterior solution, which also matches a constant density interior,
was derived in \cite{GeMa01}.

Several classes of spherically symmetric solutions of the static
gravitational field equations in the vacuum on the brane have been
obtained in~\cite{Ha03,Ma04,Ha04,Ha05}. As a possible physical
application of these solutions the behavior of the angular
velocity $v_{tg}$ of the test particles in stable circular orbits
has been considered~\cite{Ma04,Ha04,Ha05}. The general form of the
solution, together with two constants of integration, uniquely
determines the rotational velocity of the particle. In the limit
of large radial distances, and for a particular set of values of
the integration constants the angular velocity tends to a constant
value. This behavior is typical for massive particles (hydrogen
clouds) outside galaxies, and is usually explained by postulating
the existence of dark matter. The exact galactic metric, the dark
radiation, the dark pressure and the lensing in the flat rotation
curves region in the brane world scenario has been obtained in
\cite{Ha05}. It is also interesting to note that the flat rotation
curves can also be explained in $f(R)$ modified theories of
gravity, without the need of dark matter \cite{darkmatter}.

Furthermore, two families of analytic solutions of the spherically
symmetric vacuum brane world model equations (with $g_{tt}\neq
-1/g_{rr}$), parameterized by the ADM mass and a PPN parameter
$\beta $ have been obtained in \cite{cfm02}. Non-singular
black-hole solutions in the brane world model have been considered
in \cite{Da03}, by relaxing the condition of the zero scalar
curvature but retaining the null energy condition. The
four-dimensional Gauss and Codazzi equations for an arbitrary
static spherically symmetric star in a Randall--Sundrum type II
brane world have been completely solved on the brane in
\cite{Vi03}. The on-brane boundary can be used to determine the
full $5$-dimensional space-time geometry. The procedure can be
generalized to solid objects such as planets. A method to extend
into the bulk asymptotically flat static spherically symmetric
brane world metrics has been proposed in \cite{Ca03}. The exact
integration of the field equations along the fifth coordinate was
done by using the multipole ($1/r$) expansion. The results show
that the shape of the horizon of the brane black hole solutions is
very likely a flat ``pancake'' for astrophysical sources.
The general solution to the trace of the 4-dimensional Einstein
equations for static, spherically symmetric configurations has
been used as a basis for finding a general class of black hole
metrics, containing one arbitrary function $g_{tt}=A(r)$, which
vanishes at some $r=r_h>0$ (the horizon radius) in \cite{BMD03}.
Under certain reasonable restrictions, black hole metrics are
found, with or without matter. Depending on the boundary
conditions the metrics can be asymptotically flat, or have any
other prescribed asymptotic. For a review of the black hole
properties and of the lensing in the brane world models see
\cite{MaMu05}.

Now, to be viable models, the proposed models need to pass the
astrophysical and cosmological observational tests. There are
several possibilities of observationally testing the brane world
models at an astrophysical/cosmological scale, such as using the
time delay of gamma ray bursts \cite{HaE} or by using the
luminosity distance--redshift relation  for supernovae at higher
redshifts \cite{GeE}.

In addition to these possibilities, we also mention work on the gravitational lensing in brane world models \cite{lensingbrane}, on the role of the brane charge in orbital models of high-frequency quasiperiodic
oscillations observed in neutron star binary systems \cite{branecharge}, on the complete set of analytical solutions of the geodesic equation of
massive test particles in higher dimensional spacetimes which can be applied to brane world models \cite{branegeodeqs}, and on brane world corrections to the charged rotating black holes, to the perturbations in the electromagnetic potential and test particle motion around brane world black holes \cite{branepert}. The classical tests of general relativity,
namely, light deflection, time delay and perihelion shift, have
been analyzed, for gravitational theories with large
non-compactified extra-dimensions, in the framework of the
five-dimensional extension of the Kaluza-Klein theory, using an
analogue of the four-dimensional Schwarzschild metric in
\cite{Li00}. Solar system data also imposes some strong
constraints on Kaluza-Klein type theories. The existence of
extra-dimensions and of the brane world models can also be tested
via the gravitational radiation coming from primordial black
holes, with masses of the order of the lunar mass, $M \sim
10^{-7}M_{\odot }$, which might have been produced when the
temperature of the universe was around 1 TeV. If a significant
fraction of the dark halo of our galaxy consists of these lunar
mass black holes, a huge number of black hole binaries could
exist. The detection of the gravitational waves from these
binaries could confirm the existence of extra-dimensions
\cite{In03}.

It is the purpose of the present paper to consider the classical
tests (perihelion precession, light bending and radar echo delay)
of general relativity for static gravitational fields in the
framework of brane world gravity. To do this we shall adopt for
the geometry outside a compact, stellar type object (the Sun),
specific static and spherically symmetric vacuum solutions in the
context of brane worlds. As a first step in our study, we consider
the classical tests of general relativity in arbitrary spherically
symmetric spacetimes, and develop a general formalism that can be
used for any given metric. In particular, we consider the motion
of a particle (planet), and analyze the perihelion precession, and
in addition to this, by considering the motion of a photon, we
study the bending of light by massive astrophysical objects and
the radar echo delay, respectively. Existing data on light-bending
around the Sun, using long-baseline radio interferometry, ranging
to Mars using the Viking lander, and the perihelion precession of
Mercury, can all give significant and detectable Solar System
constraints, associated with the brane world vacuum solutions.
More precisely, the study of the classical general relativistic
tests, constrain the parameters of the various solutions analyzed.

This paper is organized in the following manner. In Section
\ref{SecII}, we outline, for self-completeness and
self-consistency, the field equations in brane world models. In
Sec. \ref{SecIII}, we consider the classical Solar System tests in
general relativity, namely, the perihelion shift, the light
deflection and the radar echo delay, for arbitrary spherically
symmetric spacetimes. In Sec. \ref{SecIV}, we analyze the
classical Solar System tests for the case of various brane world
vacuum solutions. We conclude and discuss our results in Sec.
\ref{SecV}.

\section{ Gravitational field equations on the brane}\label{SecII}

We start by considering a 5D spacetime (the bulk), with a
single 4D brane, on which matter is confined.
The 4D brane world $({}^{(4)}M,g_{\mu \nu })$ is
located at a hypersurface $\left(B\left( X^{A}\right) =0\right)$
in the 5D bulk spacetime $({}^{(5)}M,g_{AB})$, where the
coordinates are described by $X^{A},A=0,1,\ldots,4$. The induced 4D
coordinates on the brane are $x^{\mu },\mu =0,1,2,3$.

The action of the system is given by~\cite{SMS00}
\begin{equation}
S=S_{bulk}+S_{brane},  \label{bulk}
\end{equation}
where
\begin{equation}
S_{bulk}=\int\limits_{{}^{(5)}M}\sqrt{-{}^{(5)}g}
\left[\frac{1}{2k_{5}^{2}}{}
^{(5)}R+{}^{(5)}L_{m}+\Lambda_{5}\right] d^{5}X,
\end{equation}
and
\begin{equation}
S_{brane}=\int\limits_{{}^{(4)}M}\sqrt{-{}^{(5)}g}
\left[\frac{1}{k_{5}^{2}} K^{\pm}+L_{brane}\left( g_{\alpha \beta
},\psi \right)+ \lambda_{b}\right] d^{4}x,
\end{equation}
where $k_{5}^{2}=8\pi G_{5}$ is the 5D gravitational constant;
${}^{(5)}R$ and ${}^{(5)}L_{m}$ are the 5D scalar curvature and
the matter Lagrangian in the bulk, $L_{brane}\left( g_{\alpha
\beta },\psi\right)$ is the 4D Lagrangian, which is given by a
generic functional of the brane metric $ g_{\alpha \beta }$ and of
the matter fields $\psi$; $K^{\pm}$ is the trace of the extrinsic
curvature on either side of the brane; and $\Lambda_{5}$ and
$\lambda_b$ (the constant brane tension) are the negative vacuum
energy densities in the bulk and on the brane, respectively.

The energy-momentum tensor of bulk matter fields is defined as
\begin{equation}
{}^{(5)}\tilde{T}_{IJ}\equiv - 2\frac{\delta {}^{(5)}L_{m}} {\delta
{}^{(5)}g^{IJ}} +{}^{(5)}g_{IJ} {}^{(5)}L_{m}\,,
\end{equation}
while $T_{\mu \nu }$ is the energy-momentum tensor localized on
the brane and is given by
\begin{equation}
T_{\mu \nu }\equiv -2\frac{\delta L_{brane}}{\delta g^{\mu \nu
}}+g_{\mu \nu }\text{ }L_{brane}.
\end{equation}

Thus, the Einstein field equations in the bulk are given
by~\cite{SMS00}
\begin{equation}
{}^{(5)}G_{IJ}=k_{5}^{2} {}^{(5)}T_{IJ}\,,
\end{equation}
with
\begin{equation}
{}^{(5)}T_{IJ} =-\Lambda _{5}
{}^{(5)}g_{IJ}+\delta(B)\left[-\lambda_{b}
{}^{(5)}g_{IJ}+T_{IJ}\right] \,.
\end{equation}
The delta function $\delta \left( B\right) $ denotes the
localization of brane contribution. In the 5D spacetime a brane is
a fixed point of the $ Z_{2}$ symmetry. The basic equations on the
brane are obtained by projections onto the brane world. The
induced 4D metric is $ g_{IJ}={}^{(5)}g_{IJ}-n_{I}n_{J}$, where
$n_{I}$ is the space-like unit vector field normal to the brane
hypersurface ${}^{(4)}M$. In the following we assume
${}^{(5)}L_{m}=0$.

Assuming a metric of the form
$ds^{2}=(n_{I}n_{J}+g_{IJ})dx^{I}dx^{J}$, with $n_{I}dx^{I}=d\chi
$ the unit normal to the $\chi = \mathrm{constant}$ hypersurfaces
and $g_{IJ}$ the induced metric on $\chi = \mathrm{constant}$
hypersurfaces, the effective 4D gravitational equation on the
brane (the Gauss equation), takes the form~\cite{SMS00}:
\begin{equation}
G_{\mu \nu }=-\Lambda g_{\mu \nu }+k_{4}^{2}T_{\mu \nu} + k_{5}^{4}S_{\mu
\nu}-E_{\mu \nu },  \label{Ein}
\end{equation}
where $S_{\mu \nu }$ is the local quadratic energy-momentum correction
\begin{equation}
S_{\mu \nu }=\frac{1}{12}TT_{\mu \nu }-\frac{1}{4}T_{\mu}{}^{\alpha }
T_{\nu
\alpha }+\frac{1}{24}g_{\mu \nu }\left( 3T^{\alpha \beta } T_{\alpha
\beta}-T^{2}\right) ,
\end{equation}
and $E_{\mu \nu }$ is the non-local effect from the free bulk
gravitational field, the transmitted projection of the bulk Weyl
tensor $C_{IAJB}$, $ E_{IJ}=C_{IAJB}n^{A}n^{B}$, with the property
$E_{IJ}\rightarrow E_{\mu \nu }\delta _{I}^{\mu }\delta _{J}^{\nu
}\quad $as$\quad \chi \rightarrow 0$. We have also denoted
$k_{4}^{2}=8\pi G$, with $G$ the usual 4D gravitational constant.
The 4D cosmological constant, $\Lambda $, and the 4D coupling
constant, $ k_{4}$, are related by $\Lambda =k_{5}^{2}(\Lambda
_{5}+k_{5}^{2}\lambda _{b}^{2}/6)/2$ and
$k_{4}^{2}=k_{5}^{4}\lambda _{b}/6$, respectively. In the limit
$\lambda_{b}^{-1}\rightarrow 0$ we recover standard general
relativity \cite{SMS00}.

The Einstein equation in the bulk and the Codazzi equation also
imply the conservation of the energy-momentum tensor of the matter
on the brane, $ D_{\nu }T_{\mu }{}^{\nu }=0$, where $D_{\nu }$
denotes the brane covariant derivative. Moreover, from the
contracted Bianchi identities on the brane it follows that the
projected Weyl tensor obeys the constraint $D_{\nu }E_{\mu
}{}^{\nu }=k_{5}^{4}D_{\nu }S_{\mu }{}^{\nu }$.

The symmetry properties of $E_{\mu \nu }$ imply that in general we
can decompose it irreducibly with respect to a chosen $4$-velocity
field $ u^{\mu} $ as
\begin{equation}
E_{\mu \nu }=-k^{4}\left[ U\left( u_{\mu }u_{\nu} +\frac{1}{3}h_{\mu \nu
}\right) +P_{\mu \nu }+2Q_{(\mu }u_{\nu)}\right] ,  \label{WT}
\end{equation}
where $k=k_{5}/k_{4}$, $h_{\mu \nu }=g_{\mu \nu }+u_{\mu }u_{\nu
}$ projects orthogonal to $u^{\mu }$, the ``dark radiation'' term
$U=-k^{4}E_{\mu \nu }u^{\mu }u^{\nu }$ is a scalar, $Q_{\mu
}=k^{4}h_{\mu }^{\alpha }E_{\alpha \beta }$ is a spatial vector
and $P_{\mu \nu }=-k^{4}\left[ h_{(\mu }\text{ } ^{\alpha }h_{\nu
)}\text{ }^{\beta }-\frac{1}{3}h_{\mu \nu }h^{\alpha \beta }
\right] E_{\alpha \beta }$ is a spatial, symmetric and trace-free
tensor.

In the case of the vacuum state we have $\rho =p=0$, $T_{\mu \nu
}\equiv 0$, and consequently $S_{\mu \nu }=0$. Therefore the field
equation describing a static brane takes the form
\begin{equation}
R_{\mu \nu }=-E_{\mu \nu }+\Lambda g_{\mu \nu },
\end{equation}
with the trace $R$ of the Ricci tensor $R_{\mu \nu }$ satisfying the
condition $R=R_{\mu }^{\mu }=4\Lambda $.

In the vacuum case $E_{\mu \nu }$ satisfies the constraint $D_{\nu
}E_{\mu }{}^{\nu }=0$. In an inertial frame at any point on the
brane we have $ u^{\mu }=\delta _{0}^{\mu}$ and
$h_{\mu\nu}=\mathrm{diag}(0,1,1,1)$. In a static vacuum
$Q_{\mu}=0$ and the constraint for $E_{\mu\nu}$ takes the form~
\cite{GeMa01}
\begin{equation}
\frac{1}{3}D_{\mu }U+\frac{4}{3}UA_{\mu }+D^{\nu }P_{\mu \nu }+A^{\nu}
P_{\mu \nu }=0,
\end{equation}
where $A_{\mu }=u^{\nu }D_{\nu }u_{\mu }$ is the 4-acceleration.
In the static spherically symmetric case we may choose $A_{\mu
}=A(r)r_{\mu }$ and $ P_{\mu \nu }=P(r)\left( r_{\mu }r_{\nu
}-\frac{1}{3}h_{\mu \nu }\right) $, where $A(r)$ and $P(r)$ (the
``dark pressure'' although the name dark anisotropic stress might
be more appropriate) are some scalar functions of the radial
distance~$r$, and~$r_{\mu }$ is a unit radial vector~\cite{Da00}.

In order to obtain results which are relevant to the Solar System
dynamics, in the following we will restrict our study to the
static and spherically symmetric metric given by
\begin{equation}
ds^{2}=-e^{\nu (r)}dt^{2}+e^{\lambda (r)}dr^{2}+r^{2}d\Omega ^{2},
\label{metr1}
\end{equation}
where $d\Omega ^{2}=d\theta ^{2}+\sin ^{2}\theta d\phi ^{2}$. Here
$\theta $ and $\phi $ are the standard coordinates on the sphere,
$t\in R$ and $r$ ranges over an open interval $\left( r_{\min
},r_{\max }\right) $ so that $ -\infty \leq r_{\min }\leq r_{\max
}\leq \infty $. We also assume that the functions $\nu (r)$ and
$\lambda (r)$ are strictly positive and (at least piecewise)
differentiable on the interval $\left( r_{\min },r_{\max }\right)
$.

Then the gravitational field equations and the effective
energy-momentum tensor conservation equation in the vacuum take
the form \cite {Ha03,Ma04,Ha04}
\begin{equation}
-e^{-\lambda }\left( \frac{1}{r^{2}}-\frac{\lambda ^{\prime
}}{r}\right) + \frac{1}{r^{2}}=\frac{48\pi G}{k_{4}^{4}\lambda
_{b}}U+\Lambda ,  \label{f1}
\end{equation}
\begin{equation}
e^{-\lambda }\left( \frac{\nu ^{\prime
}}{r}+\frac{1}{r^{2}}\right) -\frac{1 }{r^{2}}=\frac{16\pi
G}{k_{4}^{4}\lambda _{b}}\left( U+2P\right) -\Lambda , \label{f2}
\end{equation}
\begin{equation}
e^{-\lambda }\left( \nu ^{\prime \prime }+\frac{\nu ^{\prime
2}}{2}+\frac{ \nu ^{\prime }-\lambda ^{\prime }}{r}-\frac{\nu
^{\prime }\lambda ^{\prime } }{2}\right) =\frac{32\pi
G}{k_{4}^{4}\lambda _{b}}\left( U-P\right) -2\Lambda ,  \label{f3}
\end{equation}
\begin{equation}
\nu ^{\prime }=-\frac{U^{\prime }+2P^{\prime }}{2U+P}-\frac{6P}{r\left(
2U+P\right) },  \label{f4}
\end{equation}
where we denoted $^{\prime }=d/dr$.

Equation (\ref{f1}) can immediately be integrated to give
\begin{equation}
e^{-\lambda }=1-\frac{C_{1}}{r}-\frac{2GM_U\left( r\right)
}{r}-\frac{ \Lambda }{3}r^2,  \label{m1}
\end{equation}
where $C_{1}$ is an arbitrary constant of integration, and we
denoted
\begin{equation}  \label{mu}
M_U\left( r\right) =\frac{24\pi }{k_4^{4}\lambda _b}\int _0^r
r^{2}U\left( r\right) dr.
\end{equation}

The function $M_U$ is the gravitational mass corresponding to the
dark radiation term (the dark mass). By substituting $\nu ^{\prime }$ given by Eq. (\ref{f4})
into Eq. (\ref{f2}) and with the use of Eq. (\ref{m1}) we obtain the
following system of differential equations satisfied by the dark
radiation term $U$, the dark pressure $P$ and the dark mass $M_U$,
describing the vacuum gravitational field, exterior to a massive
body, in the brane world model:
\begin{eqnarray}
\frac{dU}{dr}&=&-\frac{\left( 2U+P\right) \left[ 2GM+M_{U}+\alpha
\left( U+2P\right) r^{3}\right] -\frac{2}{3}\Lambda
r^{2}}{r^{2}\left(1-\frac{2GM}{ r}-\frac{M_{U}}{r}-\frac{\Lambda
}{3}r^{2}\right) }
    \nonumber   \\
&&-2\frac{dP}{dr}-\frac{6P}{ r}\,,  \label{e1}
\end{eqnarray}
and
\begin{equation}
\frac{dM_{U}}{dr}=3\alpha r^{2}U\,,  \label{e2}
\end{equation}
respectively.

The system of equations (\ref{e1}) and (\ref{e2}) can be
transformed to an autonomous system of differential equations by
means of the transformations
\begin{eqnarray}
&&q=\frac{2GM}{r}+\frac{M_{U}}{r}+\frac{\Lambda }{3}r^{2},\quad
\mu =3\alpha r^{2}U+3r^{2}\Lambda  \,,
  \nonumber  \\
&&p=3\alpha r^{2}P-3r^{2}\Lambda ,\quad \theta =\ln r\,.
\label{trans}
\end{eqnarray}
We may call $\mu $ and $p$ the ``reduced'' dark radiation and
pressure, respectively.

With the use of the new variables given by Eqs. (\ref{trans}),
Eqs. (\ref{e1}) and (\ref{e2}) become
\begin{equation}
\frac{dq}{d\theta }=\mu -q\,,  \label{aut1}
\end{equation}
and
\begin{equation}
\frac{d\mu }{d\theta }=-\frac{\left( 2\mu +p\right) \left[
q+\frac{1}{3} \left( \mu +2p\right) \right]
}{1-q}-2\frac{dp}{d\theta }+2\mu -2p\,, \label{aut2}
\end{equation}
respectively

Equations (\ref{e1}) and (\ref{e2}), or equivalently, (\ref{aut1})
and (\ref {aut2}) may be called the structure equations of the
vacuum on the brane. In order to close this system an ``equation
of state'' relating the reduced dark radiation and the dark
pressure terms is needed. Generally, this equation of state is
given in the form $P=P(U)$. In the new variables the functional
relation between dark radiation and dark pressure takes the form
$p=3\alpha \exp \left( 2\theta \right) P\left[ \mu /3\alpha \exp
\left( 2\theta \right) \right]$. We consider several specific
solutions below in the context of the Solar System tests.

\section{Classical tests of general relativity in arbitrary spherically
symmetric static space-times}\label{SecIII}

At the level of the Solar System there are three fundamental
tests, which can provide important observational evidence for
general relativity and its generalization, and for alternative
theories of gravitation in flat space. These tests are the
perihelion precession of Mercury, the deflection of light by the
Sun and the radar echo delay observations. These tests have been
used to successfully test the Schwarzschild solution of general
relativity. In order to constrain the brane world models at the
level of the Solar System, we first have to study these effects in
spherically symmetric space-times with arbitrary metrics. In this
Section, we develop a formalism that can be used for any given
metric. This useful formalism was first used in the
Solar System tests applied to a specific vacuum solution in
Ho\v{r}ava-Lifshitz gravity \cite{Harko:2009qr}.

\subsection{The perihelion precession}

The motion of a test particle in the gravitational field on the
brane in the metric given by Eq. (\ref{metr1}) can be derived from
the variational principle
\begin{equation}
\delta \int \sqrt{e^{\nu }c^{2}\dot{t}^{2}-e^{\lambda }\dot{r}
^{2}-r^{2}\left( \dot{\theta}^{2}+\sin ^{2}\theta
\dot{\phi}^{2}\right) } ds=0,  \label{var}
\end{equation}
where the dot denotes $d/ds$. It may be verified that the orbit is
planar, and hence we can set $\theta =\pi /2$ without any loss of
generality. Therefore we will use $\phi $ as the angular
coordinate. Since neither $t$ nor $\phi $ appear explicitly in Eq.
(\ref{var}), their conjugate momenta yield constants of motion,
\begin{equation}
e^{\nu }c^{2}\dot{t}=E={\rm constant}, \qquad
r^{2}\dot{\phi}=L={\rm constant}. \label{consts}
\end{equation}
The constant $E$ is related to energy conservation while the constant
$L$ is related to angular momentum conservation.

The line element Eq. (\ref{metr1}) provides the following equation
of motion for $r$
\begin{equation}
\dot{r}^{2}+e^{-\lambda }r^{2}\dot{\phi}^{2}=e^{-\lambda }\left(
e^{\nu }c^{2}\dot{t}^{2}-1\right) \,.
  \label{constants}
\end{equation}

Substitution of $\dot{t}$ and $\dot{\phi}$ from Eqs. (\ref{consts})
yields the following relationship
\begin{equation}
\dot{r}^{2}+e^{-\lambda }\frac{L^{2}}{r^{2}}=e^{-\lambda
}\left(\frac{E^{2} }{c^{2}}e^{-\nu }-1\right) .  \label{inter1}
\end{equation}

The change of variable $r=1/u$ and the substitution
$d/ds=Lu^{2}d/d\phi $ transforms Eq. (\ref{inter1}) into the form
\begin{equation}
\left( \frac{du}{d\phi }\right) ^{2}+e^{-\lambda
}u^{2}=\frac{1}{L^{2}} e^{-\lambda }\left(
\frac{E^{2}}{c^{2}}e^{-\nu }-1\right) .
\end{equation}

By formally representing $e^{-\lambda }=1-f(u)$, we obtain
\begin{equation}
\left( \frac{du}{d\phi }\right)
^{2}+u^{2}=f(u)u^{2}+\frac{E^{2}}{c^{2}L^{2}} e^{-\nu -\lambda
}-\frac{1}{L^{2}}e^{-\lambda }\equiv G(u). \label{ueq_basic}
\end{equation}

By taking the derivative of the previous equation with respect to
$\phi $ we find
\begin{equation}
\frac{d^{2}u}{d\phi ^{2}}+u=F(u),  \label{inter2}
\end{equation}
where
\begin{equation}
F(u)=\frac{1}{2}\frac{dG(u)}{du}.
\end{equation}

A circular orbit $u=u_{0}$ is given by the root of the equation $
u_{0}=F\left( u_{0}\right) $. Any deviation $\delta =u-u_{0}$ from
a circular orbit must satisfy the equation
\begin{equation}
\frac{d^{2}\delta }{d\phi ^{2}}+\left[ 1-\left( \frac{dF}{du}\right)
_{u=u_{0}}\right] \delta =O\left( \delta ^{2}\right) ,
\end{equation}
which is obtained by substituting $u=u_{0}+\delta $ in Eq.
(\ref{inter2}). Therefore, to first order in $\delta $, the
trajectory is given by
\begin{equation}
\delta =\delta _{0}\cos \left( \sqrt{1-\left(
\frac{dF}{du}\right)_{u=u_{0}} }\phi +\beta \right) ,
\end{equation}
where $\delta _{0}$ and $\beta $ are constants of integration. The
angles of the perihelia of the orbit are the angles for which $r$
is minimum and hence $u$ or $\delta $ is maximum. Therefore, the
variation of the orbital angle from one perihelion to the next is
\begin{equation}
\phi =\frac{2\pi }{\sqrt{1-\left( \frac{dF}{du}\right)
_{u=u_{0}}}}=\frac{ 2\pi }{1-\sigma }.  \label{prec}
\end{equation}

The quantity $\sigma $ defined by the above equation is called the
perihelion advance. It represents the rate of advance of the
perihelion. As the planet advances through $\phi $ radians in its
orbit, its perihelion advances through $\sigma \phi $ radians.
From Eq. (\ref{prec}), $\sigma $ is given by
\begin{equation}
\sigma =1-\sqrt{1-\left( \frac{dF}{du}\right) _{u=u_{0}}},
\end{equation}
or, for small $\left( dF/du\right) _{u=u_{0}}$, by
\begin{equation}
\sigma =\frac{1}{2}\left( \frac{dF}{du}\right) _{u=u_{0}}.
\end{equation}

For a complete rotation we have $\phi \approx 2\pi (1+\sigma)$,
and the advance of the perihelion is $\delta \phi =\phi -2\pi
\approx 2\pi \sigma $. In order to be able to perform effective
calculations of the perihelion precession we need to know the
expression of $L$ as a function of the orbit parameters. Let's
consider the motion of a planet on a Keplerian ellipse with
semi-axis $a$ and $b$, where $b=a\sqrt{1-e^{2}}$, and $e$ is the
eccentricity of the orbit. The surface area of the ellipse is $\pi
ab$. Since the elementary oriented surface area of the ellipse is
$d\vec{\sigma} =\left( \vec{r}\times d\vec{r}\right) /2$, the
areolar velocity of the planet is $\left| d\vec{\sigma}/dt\right|
=\left| \vec{r}\times d\vec{r} \right| /2=r^{2}\left( d\phi
/dt\right) /2\approx \pi a^{2}\sqrt{1-e^{2}}/T$ , where $T$ is the
period of the motion, which can be obtained from Kepler's third
law as $T^{2}=4\pi ^{2}a^{3}/GM$. In the small velocity limit $
ds\approx cdt$, and the conservation of the relativistic angular
momentum gives $r^{2}d\phi /dt=cL$. Therefore we obtain $L=2\pi
a^{2}\sqrt{1-e^{2}} /cT $ and $1/L^{2}=c^{2}/GMa\left(
1-e^{2}\right) $.

As a first application of the present formalism we consider the
precession of the perihelion of a planet in the Schwarzschild
geometry with $e^{\nu }=e^{-\lambda }=1-2GM/c^{2}r=1-\left(
2GM/c^{2}\right) u$. Hence $ f(u)=\left( 2GM/c^{2}\right) u$.
Since for this geometry $\nu +\lambda =0$, we obtain first
\begin{equation}
G(u)=\frac{2GM}{c^{2}}u^{3}+\frac{1}{L^{2}}\left(
\frac{E^{2}}{c^{2}} -1\right) +\frac{2GM}{c^{2}L^{2}}u,
\end{equation}
and then
\begin{equation}
F(u)=3\frac{GM}{c^{2}}u^{2}+\frac{GM}{c^{2}L^{2}}.
\end{equation}

The radius of the circular orbit $u_{0}$ is obtained as the
solution of the equation
\begin{equation}
u_{0}=3\frac{GM}{c^{2}}u_{0}^{2}+\frac{GM}{c^{2}L^{2}},
\end{equation}
with the only physical solution given by
\begin{equation}
u_{0}=\frac{1\pm
\sqrt{1-12G^{2}M^{2}/c^{4}L^{2}}}{6GM/c^{2}}\approx \frac{GM
}{c^{2}L^{2}}.
\end{equation}

Therefore
\begin{equation}
\delta \phi =\pi \left( \frac{dF}{du}\right) _{u=u_{0}}=\frac{6\pi
GM}{ c^{2}a\left( 1-e^{2}\right) },
\end{equation}
which is the standard general relativistic result.

\subsection{The deflection of light}

In the absence of external forces a photon follows a null
geodesic, $ ds^{2}=0 $. The affine parameter along the photon's
path can be taken as an arbitrary quantity, and we denote again by
a dot the derivatives with respect to the arbitrary affine
parameter. There are two constants of motion, the energy $E$ and
the angular momentum $L$, given by Eqs. (\ref{consts}).

The equation of motion of the photon is
\begin{equation}
\dot{r}^{2}+e^{-\lambda }r^{2}\dot{\phi}^{2}=e^{\nu -\lambda
}c^{2}\dot{t} ^{2},
\end{equation}
which, with the use of the constants of motion can be written as
\begin{equation}
\dot{r}^{2}+e^{-\lambda}\frac{L^{2}}{r^{2}}=\frac{E^{2}}{c^{2}}e^{-\nu
-\lambda }.
\end{equation}

The change of variable $r=1/u$ and the use of the conservation
equations to eliminate the derivative with respect to the affine
parameter leads to
\begin{equation}
\left( \frac{du}{d\phi }\right)
^{2}+u^{2}=f(u)u^{2}+\frac{1}{c^{2}}\frac{ E^{2}}{L^{2}}e^{-\nu
-\lambda }\equiv P(u)\,. \label{P_eq_basic}
\end{equation}
By taking the derivative of the previous equation with respect to
$\phi $ we find
\begin{equation}
\frac{d^{2}u}{d\phi ^{2}}+u=Q(u),
\label{phot}
\end{equation}
where
\begin{equation}
Q(u)=\frac{1}{2}\frac{dP(u)}{du}.
\end{equation}

In the lowest approximation, in which the term of the right hand
side of the equation (\ref{phot}) is neglected, the solution is a
straight line,
\begin{equation}
u=\frac{\cos \phi }{R},  \label{u0}
\end{equation}
where $R$ is the distance of the closest approach to the mass. In
the next approximation Eq. (\ref{u0}) is used on the right-hand
side of Eq. (\ref {phot}), to give a second order linear
inhomogeneous equation of the form
\begin{equation}
\frac{d^{2}u}{d\phi ^{2}}+u=Q\left( \frac{\cos \phi }{R}\right) .
\label{uQ}
\end{equation}
with a general solution given by $u=u\left( \phi \right) $. The
light ray comes in from infinity at the asymptotic angle $\phi
=-\left( \pi /2+\varepsilon \right) $ and goes out to infinity at
an asymptotic angle $ \phi =\pi /2+\varepsilon $. The angle
$\varepsilon $ is obtained as a solution of the equation $u\left(
\pi /2+\varepsilon \right) =0$, and the total deflection angle of
the light ray is $\delta =2\varepsilon $.

In the case of the Schwarzschild metric we have $\nu +\lambda =0$
and  $ f(u)=\left( 2GM/c^{2}\right) u$, which gives $P(u)=\left(
2GM/c^{2}\right) u^{3}$ and $Q(u)=\left( 3GM/c^{2}\right) u^{2}$,
respectively. In the lowest approximation order from Eqs.
(\ref{u0}) and (\ref{uQ}) we obtain the second order linear
equation
\begin{equation}
\frac{d^{2}u}{d\phi ^{2}}+u=\frac{3GM}{c^{2}R^{2}}\cos ^{2}\phi
=\frac{3GM}{ 2c^{2}R^{2}}\left( 1+\cos 2\phi \right) ,
\end{equation}
with the general solution given by
\begin{equation}
u=\frac{\cos \phi }{R}+\frac{3GM}{2c^{2}R^{2}}\left(
1-\frac{1}{3}\cos 2\phi \right) .  \label{ulight}
\end{equation}

By substituting $\phi =\pi /2+\varepsilon $, $u=0$ into Eq.
(\ref{ulight}) we obtain
\begin{equation}
\varepsilon =\frac{2GM}{c^{2}R},
\end{equation}
where we have used the relations $\cos \left( \pi /2+\varepsilon
\right) =\allowbreak -\sin \varepsilon $, $\cos \left( \pi
+2\varepsilon \right) =\allowbreak -\cos 2\varepsilon $, $\sin
\varepsilon \approx \varepsilon $ and $\cos 2\varepsilon \approx
1$. The total deflection angle in Schwarzschild geometry is
$\delta =2\varepsilon =4GM/c^{2}R$.

\subsection{Radar echo delay}

A third Solar System test of general relativity is the radar echo
delay \cite{Sh}. The idea of this test is to measure the time
required for radar signals to travel to an inner planet or
satellite in two circumstances: a) when the signal passes very
near the Sun and b) when the ray does not go near the Sun. The
time of travel of light between two planets, situated far away
from the Sun, is given by
\begin{equation}
T_{0}=\int_{-l_{1}}^{l_{2}}\frac{dy}{c},
\end{equation}
where $l_{1}$ and $l_{2}$ are the distances of the planets to the
Sun. If the light travels close to the Sun, the time travel is
\begin{equation}
T=\int_{-l_{1}}^{l_{2}}\frac{dy}{v}=
\frac{1}{c}\int_{-l_{1}}^{l_{2}}e^{\left[
\lambda (r)-\nu (r)\right]/2}dy,
\end{equation}
where $v=ce^{\left(\nu -\lambda \right)/2}$ is the speed of light
in the presence of the gravitational field. The time difference is
\begin{equation}
\delta T=T-T_0=\frac{1}{c}\int_{-l_{1}}^{l_{2}}\left\{
e^{\left[\lambda (r)-\nu (r)\right]/2}-1\right\} dy.
\end{equation}

Since $r=\sqrt{y^{2}+R^{2}},$ we have
\begin{equation}
\delta T=\frac{1}{c}\int_{-l_{1}}^{l_{2}}\left\{ e^{\left[\lambda
\left( \sqrt{ y^{2}+R^{2}}\right) -\nu \left(
\sqrt{y^{2}+R^{2}}\right) \right]/2}-1\right\} dy.
\label{delay_eq}
\end{equation}

In the case of the Schwarzschild metric $\lambda =-\nu $, $\exp
\left( \lambda /2-\nu /2\right) =\exp \left( -\nu \right) =\left(
1-2GM/c^{2}r\right) ^{-1}\approx 1+2GM/c^{2}r$, and therefore
\begin{equation}
\delta
T=\frac{2GM}{c^{3}}\int_{-l_{1}}^{l_{2}}\frac{dy}{\sqrt{y^{2}+R^{2}}}=
\frac{2GM}{c^{3}}\ln \frac{\sqrt{R^{2}+l_{2}^{2}}+l_{2}}{\sqrt{
R^{2}+l_{1}^{2}}-l_{1}}.
\end{equation}

Since
\begin{equation}
\ln
\frac{\sqrt{R^{2}+l_{2}^{2}}+l_{2}}{\sqrt{R^{2}+l_{1}^{2}}-l_{1}}
\approx\ln \frac{4l_{1}l_{2}}{R^{2}},
\end{equation}
where we have used the conditions $R^2/l_{1}^{2}\ll 1$ and
$R^2/l_{2}^{2}\ll 1$, respectively, the time delay is given by
\begin{equation}
\delta T=\frac{2GM}{c^{3}}\ln \frac{4l_{1}l_{2}}{R^{2}}.
\end{equation}

\section{Solar system tests of specific brane world models}
\label{SecIV}

The braneworld description of our universe entails a large extra
dimension and a fundamental scale of gravity that might be lower
by several orders of magnitude compared to the Planck scale
\cite{RS99a}. It is known that the Einstein field equations in
five dimensions admit more general spherically symmetric black
holes on the brane than four-dimensional general relativity. Hence
an interesting consequence of the brane world scenario is in the
nature of spherically symmetric vacuum solutions to the brane
gravitational field equations, which could represent black holes
with properties quite distinct as compared to ordinary black holes
in four dimensions. Such black holes are likely to have very
diverse cosmological and astrophysical signatures. In the present
Section we consider the Solar System test properties of several
brane world black holes, which have been obtained by solving the
vacuum gravitational field equations. There are many black hole
type solutions on the brane, and in the following we analyze three
particular examples. We will assume, for astrophysical constants,
the values \cite{Sh}:
\bea
M_{\odot}&=&1.989\times 10^{30}~{\rm kg}, \nn \\
R_{\odot}&=&6.955 \times 10^8~ {\rm m} \nn \\
c&=&2.998\times 10^{8}~{\rm m/s}, \nn \\
G&=&6.67\times 10^{-11}~{\rm m^{3}kg^{-1}s^{-2}}, \nn \\
a&=&57.91\times 10^{9}~{\rm m}, \nn \\
e&=&~0.205615
\label{astr_const}
\eea

\subsection{The DMPR brane world vacuum solution}

The first brane solution we consider is a solution of the vacuum
field equations, obtained by Dadhich, Maartens, Papadopoulos and
Rezania in \cite {Da00}, which represent the simplest
generalization of the Schwarzschild solution of general
relativity. We call this type of brane black hole as the DMPR
black hole. The Solar System tests for the DMPR solutions were
extensively analyzed in \cite{Boehmer:2008zh}, but we use the
general and novel formalism developed above as a consistency
check. For this solution the metric tensor components are given by
\begin{equation}
e^{\nu }=e^{-\lambda }=1-\frac{2m}{r}+\frac{Q}{r^{2}},
\end{equation}
where $Q$ is the so-called tidal charge parameter. In the limit $
Q\rightarrow 0$ we recover the usual general relativistic case. In
terms of the general equations discussed in section II, this class
of brane world spherical solution is characterized by a equation of
state relating dark energy and pressure: $P = -2U$. The metric is
asymptotically flat, with $\lim_{r\rightarrow \infty }\exp {(\nu
)} =\lim_{r\rightarrow \infty }\exp {(\lambda )}=1$. There are two
horizons, given by
\begin{equation}
r_{h}^{\pm}=m\pm \sqrt{m^{2}-Q}.
\end{equation}

Both horizons lie inside the Schwarzschild horizon $r_{s}=2m$,
$0\leq r_{h}^{-}\leq r_{h}^{+}\leq r_{s}$. In the brane world
models there is also the possibility of a negative $Q<0$, which
leads to only one horizon $r_{h+}$ lying outside the Schwarzschild
horizon,
\begin{equation}
r_{h+}=m+\sqrt{m^{2}+Q}>r_{s}.
\end{equation}

In this case the horizon has a greater area than its general
relativistic counterpart, so that bulk effects act to increase the
entropy and decrease the temperature, and to strengthen the
gravitational field outside the black hole.

\subsubsection{Perihelion precession}

For the DMPR black hole we have the following relevant functions:
\begin{equation}
f(u)=2mu-Qu^{2} \,,
\end{equation}
\begin{equation}
G(u)=2mu^{3}-Qu^{4}+\frac{E^{2}}{c^2L^2}-\frac{1}{L^2}
+\frac{2mu}{L^2}-\frac{Qu^{2}}{L^2} \,,
\end{equation}
and
\begin{equation}
F(u)=3mu^{2}-2Qu^{3}+\frac{m}{L^2}-\frac{Qu}{L^2} \,,
\end{equation}
respectively.

$u_{0}$ can be obtained as solution of the algebraic equation
\begin{equation}
3mu_{0}^{2}-u_{0}+\frac{m}{L^{2}}=2Qu_{0}^{3}+\frac{Q}{L^{2}}u_{0}\,,
\label{u0DMPR}
\end{equation}
which to first order may be approximated to $u_0\approx
GM/(c^2L^2)$, assuming that $Q/L^2\ll 1$. We have used the following relationship $m = GM/c^2$. Thus, $\delta \phi$
takes the form
\begin{equation}
\delta \phi=\frac{6\pi GM}{c^2a(1-e^2)}-\frac{c^2\pi
Q}{GMa(1-e^2)}  \,,\label{delsol}
\end{equation}
which is consistent with the result outlined in
\cite{Boehmer:2008zh}. The first term in Eq. (\ref{delsol}) is the
well-known general relativistic correction term for the perihelion
precession, while the second term gives the correction due to the
non-local effects arising from the Weyl curvature in the bulk.

The observed value of the perihelion precession of the planet
Mercury is $ \delta \varphi _{Obs}=43.11\pm 0.21$ arcsec per
century \cite{Sh}. The general relativistic formula for the
precession, $\delta \varphi _{GR}=6\pi
GM/c^{2}a\left(1-e^{2}\right)$, gives, using numerical values
given in (\ref{astr_const}), $\delta \varphi _{GR}=42.94$ arcsec
per century. Therefore, the difference $\Delta \varphi =\delta
\varphi _{Obs}-\delta \varphi _{GR}=0.17\pm 0.21$ arcsec per
century can be attributed to other effects. By assuming that
$\Delta \varphi $ is entirely due to the modifications of the
general relativistic Schwarzschild geometry as a result of the
five dimensional bulk effects, the observational results impose
the following general constraint on the bulk tidal parameter $Q$:
\begin{equation}
\left| Q\right| \leq \frac{GM_{\odot }a\left( 1-e^{2}\right) }{\pi
c^{2}} \Delta \varphi \,.  \label{Qobs}
\end{equation}
With the use of the observational data for Mercury, Eq.
(\ref{Qobs}) gives $ \left| Q\right| \leq (5.17 \pm 6.39)\times 10^{4}$
m$^{2}$, or in the natural system of units, with $c=\hbar =G=1$,
$\left| Q\right| \leq (1.32 \pm 1.63) \times 10^{30}$ MeV$^{-2}$.

\subsubsection{Deflection of light}

We have the following functions
\begin{equation}
P(u)=2mu^3-Qu^4+\frac{E^2}{c^2L^2},
\end{equation}
and
\begin{equation}
Q(u)=3mu^2-2Qu^3\,.
\end{equation}

Solving the differential equation
\begin{equation}
\frac{d^2u}{d\phi^2}+u=\frac{3m}{R^2}\cos^2\phi
-\frac{2Q}{R^3}\cos^3\phi\,,
\end{equation}
provides the following exact solution
\begin{eqnarray}
u(\phi)&=&\frac{\cos\phi}{R}+\frac{3m}{2R^2}
\left(1-\frac{1}{3}\cos 2\phi\right)
   \nonumber  \\
&&-\frac{Q}{4R^3}\left(\frac{9}{4}\cos\phi -\frac{1}{4}\cos 3\phi
+3\phi\sin\phi \right) \,.
     \label{DMPRulight}
\end{eqnarray}

By substituting $\phi =\pi /2+\varepsilon $, $u=0$ into Eq.
(\ref{DMPRulight}), yields the following total deflection angle
\begin{equation}
\delta \phi=2\varepsilon=\frac{4GM}{c^2R}-\frac{3\pi Q}{4R^2}=\frac{4GM}{c^2R}
\left(1-\frac{3\pi Qc^2}{16GMR}\right)  \,,
\end{equation}
which is in agreement with the results obtained in
\cite{Gergely:2009xg}.

The best available data on light deflection by the Sun come from
long baseline radio interferometry \cite{all2}, which gives
$\delta \phi _{LD}=\delta \phi _{LD}^{(GR)}\left(
1+\Delta_{LD}\right) $, with $\Delta _{LD}\leq 0.0002 \pm 0.0008$,
where $\delta \phi_{LD}^{(GR)}=1.7510 $ arcsec. Therefore light
deflection constrains the tidal parameter $Q$ as
$\left|Q\right|\leq 16GMR\Delta _{LD}/3\pi c^2 $. By taking for
$R$ the value of the radius of the Sun in (\ref{astr_const}), the
light deflection gives the constraint $\left|Q\right|\leq (6.97
\pm 27.88) \times 10^{8}$ m$^2$, or, in natural units,
$\left|Q\right|\leq (1.78 \pm 7.11) \times 10^{33}$ MeV$^{-2}$.

\subsubsection{Radar echo delay}

The delay can be evaluated from the integral in Eq.
(\ref{delay_eq}), with $\lambda=-\nu $, $\exp \left( \lambda
/2-\nu /2\right) =\exp \left(\lambda \right)$, so that
\be \exp \left(\lambda \right) =\left(
1-\frac{2GM}{c^2r}+\frac{Q}{r^2}\right) ^{-1}\approx \left(
1+\frac{2GM}{c^2r}-\frac{Q}{r^2}\right) \,,\ee
so that the time delay, given by Eq.~(\ref{delay_eq}), is readily
integrated to yield
\begin{eqnarray}
\d T &=& \frac{2GM}{c^3}\ln
\left(\frac{\sqrt{R^{2}+l_{2}^{2}}+l_{2}}{\sqrt{R^{2}
+l_{1}^{2}}-l_{1}}\right)
   \nonumber  \\
&&-\frac{Q}{cR}\left[\arctan\left(\frac{l_2}{R}\right)
+\arctan\left(\frac{l_1}{R}\right)\right] \,.
\end{eqnarray}
Using the approximations $R^2/l_{1}^{2}\ll 1$ and
$R^2/l_{2}^{2}\ll 1$, the above expression reduces to
\be \d T
\approx\frac{2GM}{c^3}\ln\left(\frac{4l_1l_2}{R^2}\right)-\frac{\pi
Q}{cR} \,. \ee

Recently the measurements of the frequency shift of radio photons
to and from the Cassini spacecraft as they passed near the Sun
have greatly improved the observational constraints on the radio
echo delay. For the time delay of the signals emitted on Earth,
and which graze the Sun, one obtains $\Delta t_{RD}=\Delta
t_{RD}^{(GR)}\left( 1+\Delta _{RD}\right) $, with $\Delta
_{RD}\simeq (1.1 \pm 1.2) \times 10^{-5}$ \cite{Reasemberg}.
Therefore radar echo delay constrains the tidal charge of the DMPR
brane world black hole as $\left|Q\right|\leq 2GMR\Delta
_{RD}\ln\left(4l_1l_2/R^2\right)/\pi c^2\approx (1.83 \pm 1.99)
\times 10^{8}$ m$^2$. In natural units we have $\left|Q\right|\leq
(4.66 \pm 5.08 )\times 10^{33}$ MeV$^{-2}$.

\subsection{The CFM solution}

Two families of analytic solutions in the brane world model,
parameterized by the ADM mass and the PPN parameters $\beta$ and
$\gamma $, and which reduce to the Schwarzschild black hole for
$\beta =1$, have been found by Casadio, Fabbri and Mazzacurati in
\cite{cfm02}. We denote the corresponding brane black hole
solutions as the CFM black holes.

The first class of solutions is given by
\begin{equation}
e^{\nu }=1-\frac{2m}{r},
\end{equation}
and
\begin{equation}
e^{\lambda }=\frac{1-\frac{3m}{2r}}{\left( 1-\frac{2m}{r}\right)
\left[ 1-
\frac{3m}{2r}\left( 1+\frac{4}{9}\eta \right) \right] },
\end{equation}
respectively, where $\eta =\gamma -1=2\left( \beta -1\right) $.
This solution is characterized by an equation of state of the form
\be \frac{P}{U} = \frac{1 - \frac{3m}{4r}}{\frac{m}{3r}}
\label{CFM1eos} \ee As in the Schwarzschild case the event horizon
is located at $r=r_{h}=2m$. The solution is asymptotically flat,
that is $\lim_{r\rightarrow \infty }e^{\nu }=e^{\nu _{\infty
}}=\lim_{r\rightarrow \infty }e^{\lambda }=e^{\lambda _{\infty
}}=1$. We consider this case in the analysis outlined below as the
CFM1 solution.

The second class of solutions corresponding to brane world black
holes obtained in \cite{cfm02} has the metric tensor components
given by
\begin{equation}
e^{\nu }=\left[ \frac{\eta +\sqrt{1-\frac{2m}{r}\left( 1+\eta
\right) }}{ 1+\eta }\right] ^{2},  \label{cfm021}
\end{equation}
and
\begin{equation}
e^{\lambda }=\left[ 1-\frac{2m}{r}\left( 1+\eta \right) \right] ^{-1},
\label{cfm022}
\end{equation}
respectively. The equation of state describing this solution is $U
= 0$ The metric is asymptotically flat. In the case $\eta >0$, the
only singularity in the metric is at $r=r_{0}=2m\left(1+\eta
\right) $, where all the curvature invariants are regular.
$r=r_{0}$ is a turning point for all physical curves. For $\eta
<0$ the metric is singular at $ r=r_{h}=2m/\left( 1-\eta \right) $
and at $r_{0}$, with $r_{h}>r_{0}$, where $ r_{h} $ defines the
event horizon. We consider this case in the analysis outlined
below as the CFM2 solution.

\subsubsection{Perihelion precession}

The function $G(u)$ in Eq. (\ref{ueq_basic}) specializes, for the
CFM1 metric and CFM2 metric in
\bea G_{1}(u) &=& \left[ 1- \frac{(1-2mu)\left(1-
\frac{3}{2}\left(1+
\frac{4}{9}\eta\right)mu\right)}{(1-\frac{3}{2}mu)}\right]u^2
   \nn  \\
&&+ \frac{1}{L^2}\frac{(1-2mu)\left(1-\frac{3}{2}\left(1+
\frac{4}{9}\eta\right) mu \right)}{(1-\frac{3}{2}mu)}\times
    \nn \\
&&\times\left[\frac{E^2}{c^2}\frac{1}{(1-2mu)} - 1 \right]\,, \eea
and
\bea G_{2}(u) &=& 2m(1+\eta)u^3-\frac{1-2m(1+\eta)u}{L^2}
     \nn  \\
&&+ \frac{E^2}{c^2 L^2} (1+\eta)^2 \frac{1-2m(1+\eta)u}{\left(\eta
+ \sqrt{1-2m(1+\eta)u}\right)^2}\,,\nn \\
 \label{G_CFM} \eea
respectively.

The GR limit is obtained for $\eta =0$, which gives the expected
value of the PPN parameter $\g = 1$. The function $F(u)$ is quite
complicated, but we are only interested in the first order
corrections to the GR results. Since the terms $m/L$, $\eta$ and
$E^2/c^2 -1$ are very small, we can expand $F(u)$ up to second
order, and keep only the term of order $O(2)$ in the product of
the infinitesimals. We get, for the CFM1 and CFM2 solutions,
respectively:
\bea F_1(u) &\simeq& \frac{m}{L^2} + \frac{3(1+\frac{\eta}{3})} m
u^2\,,
 \\
F_2(u) &\simeq& \frac{m}{L^2} + \frac{3(1+\eta)} m u^2\,.
\label{F_CFM} \eea
Both function can be written as \be F_{1,2}(u) \equiv
\frac{\tilde{m}_{1,2}}{\tilde{L}_{1,2}^2} + 3
\tilde{m}_{1,2}u^2\,, \label{F_CFM_tilde} \ee where the tilded
parameters are defined as: \bea \tilde{m}_1 = (1+
\frac{\eta}{3})m;&~~~~~~&\tilde{L}^2_1 =
(1 + \frac{\eta}{3}) L^2\,, \nn \\
\tilde{m}_2 = (1 + \eta)m;&~~~~~~&\tilde{L}^2_1 = (1 + \eta + 1)
L^2\,. \nn \label{tildedpar} \eea
This allows us to use directly the result obtained in GR,
substituting the values of the tilded parameters. The resulting
perihelion precession is: \bea \delta \phi_1 \simeq \frac{6 \pi G
M}{c^2 a(1-e^2)} +
\frac{2 \pi G M}{c^2 a(1-e^2)}\eta\,,  \\
\delta \phi_2 \simeq \frac{6 \pi G M}{c^2 a(1-e^2)} + \frac{6 \pi
G M}{c^2 a(1-e^2)}\eta \,,\label{Delta_phi_CFM} \eea where we have
used the value of the Schwarzschild radius of the sun $m =
GM_{\odot}/c^2$, as mentioned above. As in the previous section, we consider the case of the planet Mercury and assume  that the discrepancy between the
predicted and the observed value is completely due to extradimension influence. We thus obtain a limit for the value of $\eta$ and consequently for $\beta$. In the CFM2 case (which is the most constraining) we have the following restrictions: \be |\eta| \leq 0.004 \pm 0.005 \label{PPNlimitsp} \ee The $\eta$
parameter for the CFM1 solution has a bound three times larger.

\subsubsection{Deflection of light}

The function $P(u)$ in Eq. (\ref{P_eq_basic}) for the CFM1 and the
CFM2 metric takes the following respective forms:
\bea P_1(u) &=& \left[ 1- \frac{(1-2mu)\left(1-\frac{3}{2}\left(1+
\frac{4}{9}\eta\right) mu\right)}{(1-\frac{3}{2}mu)}\right]u^2
   \nn  \\
&&+ \frac{E^2}{c^2 L^2}\frac{\left[1-\frac{3}{2}\left(1+
\frac{4}{9}\eta\right) mu\right]}{(1-\frac{3}{2}mu)}\,, \\
P_{2}(u) &=& 2m(1+\eta)u^3+ \frac{E^2}{c^2 L^2}(1+\eta)^2 \times
   \nn  \\
&&\times \frac{1-2m(1+\eta)u}{\left(\eta +
\sqrt{1-2m(1+\eta)u}\right)^2}\,. \label{P_CFM} \eea

The function $Q(u)$ of Eq. (\ref{phot}) can be expanded as:
\bea Q_1(u) &\simeq& -\frac{\eta}{3} \frac{E^2}{c^2}\frac{m}{L^2}
- \eta \frac{E^2}{c^2}\frac{m^2}{L^2}u
  \nn  \\
&&+ 3m\left[ 1+ \frac{\eta}{3} - \frac{3}{4}\eta
\frac{E^2}{c^2}\frac{m^2}{L^2}\right]u^2 + O^3(u) \,, \eea
and
\bea Q_2(u) &\simeq& -\eta \frac{E^2}{c^2}\frac{m}{L^2} - 3\eta
\frac{E^2}{c^2}\frac{m^2}{L^2}u
    \nn  \\
&&+ 3m\left[ 1+\eta - \frac{\eta+5}{2}
\frac{E^2}{c^2}\frac{m^2}{L^2}\right]u^2 + O^3(u) \,,\label{Q_exp}
\eea
for the CFM1 and CFM2 solutions, respectively.

As stated before, at first order, we can take, for both cases,
$Q(u) = 0$, so that the zeroth-order solution is \be u^{(0)} =
\frac{\cos(\phi)}{R} \,,\label{u_0} \ee which means the light rays
travel on straight lines. Assuming, again, that the infinitesimals
$\eta$ and $m/L$ are comparable, the first order equation is the
same obtained in standard GR, namely: \be {u^{(1)}}'' + u^{(1)} =
3 m \left( \frac{\cos \phi}{R} \right)^2 \label{light_eq_1} \ee
which gives, of course, the standard GR result \be u^{(1)} =
\frac{ m}{2R^2}\left[3-\cos(2\phi)\right]. \label{u_1} \ee

At second order Eq. (\ref{phot}) can be written as \bea
{u_1^{(2)}}'' + u^{(2)}_1 &=&
-\frac{\eta}{3}\frac{E^2}{c^2}\frac{m}{L^2}
+ 6\frac{m}{L}u^{(0)}u^{(1)} + \eta\frac{m}{L} {u^{(0)}}^2 \nn \\
&&\hspace{-2.7cm} = -\frac{\eta}{3}\frac{E^2}{c^2}\frac{m}{L^2} +
\frac{3m^2}{R^3}\cos(\phi)\left[ 2-cos(2\phi)\right]
   + m\eta\frac{\cos^2\phi}{R^2} \,,
\eea
and
\bea {u_2^{(2)}}'' + u^{(2)}_2 &=& -
\eta\frac{E^2}{c^2}\frac{m}{L^2}
+ 6\frac{m}{L}u^{(0)}u^{(1)} + 3\eta\frac{m}{L} {u^{(0)}}^2 \nn \\
&& \hspace{-2.6cm}= \eta\frac{E^2}{c^2}\frac{m}{L^2} +
\frac{3m^2}{R^3}\cos(\phi)\left[ 2-cos(2\phi)\right] +
3m\eta\frac{\cos^2\phi}{R^2}\,, \label{light_eq_2} \eea
respectively for the CFM1 and CFM2 solutions. The solutions are
provided by
\bea u_1^{(2)} &=& \frac{m^2}{16 R^3} \Bigg[ 12\cos^3(\phi)
-\frac{16}{3}\eta\frac{R}{m}\cos^2(\phi)
   \nn \\
&&\hspace{-1cm}+ 21\cos(\phi) + 60\phi\sin(\phi) \Bigg]
-\frac{\eta}{3}\frac{E^2}{c^2}\frac{m}{L^2}
+\frac{2}{3}\eta\frac{m}{R^2}\,, \eea
and
\bea u_2^{(2)} &=& \frac{m^2}{16 R^3} \Bigg[ 12\cos^3(\phi)
-16\eta\frac{R}{m}\cos^2(\phi)
   \nn \\
&&\hspace{-1cm}+ 21\cos(\phi) + 60\phi\sin(\phi) \Bigg] -
\eta\frac{E^2}{c^2}\frac{m}{L^2} + 2\eta\frac{m}{R^2}\,,
\label{u_2} \eea respectively.

The deflection angle is obtained by the equation $u(\pi/2 + \e)=0$
with $\d \phi = 2\e$. Assuming $L = RE/c$ and substituting again
$m = GM_{\odot}/c^2$, we finally find:
\bea \d \phi_1 &=&
4\frac{GM_{\odot}}{c^2R} \left( 1+\frac{15}{16}\pi
\frac{GM_{\odot}}{c^2R} + \frac{\eta}{6} \right)\,,  \\
\d \phi_2 &=& 4\frac{GM_{\odot}}{c^2R} \left( 1+\frac{15}{16}\pi
\frac{GM_{\odot}}{c^2R} + \frac{\eta}{2} \right)\,.
\label{light_delta} \eea

Comparison with data from the long baseline radio interferometry
experiments \cite{Sh}, allow us, as in the previous section, to
place an upper limit for $\eta$ (and thus for $\beta$) which is,
for the CFM2 solution: \be |\eta| \leq 0.0004 \pm 0.0016
\label{PPNlimitsl} \ee whilst, as before, the $\eta$ parameter for
the CFM1 solution has a bound three times larger.

\subsubsection{Radar echo delay}

The delay can be evaluated from the integral in Eq.
(\ref{delay_eq}). The integrand, in the case of the CMF1 and CFM2
metrics, is \bea &\exp\left( \frac{\l-\nu}{2} \right) \simeq 1 +
\left(2\frac{\eta}{3}\right) m u +
\left( 4-\frac{5}{6}\eta \right)m^2u^2 + O(u^3)\,, \nn \\
&\exp\left( \frac{\l-\nu}{2} \right) \simeq 1 + (2+\eta)m u +
\left( 4+ \frac{9}{2}\eta\right)m^2u^2 + O(u^3)\nn\,,
\label{exp_delay} \eea respectively, so that the delay is given by
\begin{eqnarray}
\d T_1 &=& 2\frac{GM}{c^3}\log\frac{4l_1l_2}{R^2} \left(1 +
\frac{\eta}{6} + 2\pi\frac{GM}{c^2R}\log^{-1}\frac{4l_1l_2}{R^2}
\right)\,, \end{eqnarray} and \bea
 \d T_2 &=& 2\frac{GM}{c^3}\log\frac{4l_1l_2}{R^2}
\left(1 + \frac{\eta}{2}
 + 2\pi\frac{GM}{c^2R}\log^{-1}\frac{4l_1l_2}{R^2} \right)\,,\nn \\
\label{CFMdelay}
\end{eqnarray}

These results are to be compared with data from the Cassini
spacecraft \cite{Reasemberg} described in the previous section. We
get a bound for $\eta$ in the CFM2 model: \be |\eta| \leq 0.000021
\pm 0.000024 \label{PPNlimitsd} \ee the $\eta$ parameter for the
CFM1 solution has again a bound three times larger.

So we can conclude that the CFM models add a correction to the
relevant test quantities which are directly proportional to the
PPN parameter that describes the solution themselves, plus
additional terms quadratic in the ratio between the Schwarzschild
radius and the typical dimension of the phenomenon.

\subsection{The BMD solution}

Several classes of brane world black hole solutions have been
obtained by Bronnikov, Melnikov and Dehnen in~\cite{BMD03} (for
short the BMD black holes). A particular class of these models has
the metric given by
\begin{equation}
  e^{\nu }=\left( 1-\frac{2G m}{c^2 r}\right) ^{2/s},\qquad
  e^{\lambda }=\left( 1-\frac{2G m}{c^2 r}\right) ^{-2},
\end{equation}
where $s\in \mathbb{N}$. The metric is asymptotically flat, and at
$r=r_{h}=2 Gm/c^2$ these solutions have a double horizon.

It is also interesting to mention that the DMPR and CFM black holes, analyzed above, are part of the families of solutions classified by Bronnikov, Melnikov and Dehnen in \cite{BMD03}.

\subsubsection{Perihelion precession}

For the BMD black hole solutions $f(u)=(4Gm/c^2)u-(2Gm/c^2)^2 u^2$
and $F(u)$ results in an involved expression. The equation
$u_0=F(u_0)$ cannot be solved exactly. We therefore expand $F(u)$
in $1/c^2$ keeping only terms up to second order, which means up
to terms $1/c^4$, and find
\begin{eqnarray}
  F(u) &\simeq& \frac{2 G m}{c^2 L^2 s} + \frac{8 G^2 m^2}{c^4 L^2 s^2}u
  - \frac{12 G^2 m^2}{c^4 L^2 s}u
   \nn \\
  &&+ \frac{6 G m}{c^2}u^2 -
  \frac{8 G^2 m^2}{c^4}u^3.
\end{eqnarray}
Therefore $u_0=F(u_0)$ now becomes a cubic equation for $u_0$. The
physical solution in up to terms of the order $(Gm/c^2)^4$ is
given by
\begin{equation}
  u_0 = \frac{2Gm}{c^2 L^2 s} \left(1+\frac{8 G^2 m^2}
  {c^4 L^2 s^2}\right).
\end{equation}
Hence, the perihelion precession is
\begin{equation}
  \delta\phi = \frac{6\pi G m}{a^2 c^2 (1-e^2)}\left(1 +
  \frac{4+6s-3s^2}{3s^2}\right).
\end{equation}
Comparison with observations yields
\begin{equation}
  -0.004 < \frac{4+6s-3s^2}{3s^2} < 0.004\,,
\end{equation}
which can be achieved by the following parameter choices
\begin{equation}
  -0.528 < s < -0.527\,,\qquad\mbox{or}\qquad 2.519 < s < 2.536\,.
\end{equation}
These inequalities place stringent constraints on the parameter
$s$, and since $s \in \mathbb{N}$, the restrictions strongly
violate Solar System constrains. Thus, we can safely conclude that
the BDM solution is incompatible with Mercury's perihelion
precession.

\subsubsection{Deflection of light}

As above, the functional form of $Q(u)$ is too complicated to
solve the resulting second order differential equation. Therefore,
we expand $Q(u)$ in $1/c^2$ and find
\begin{eqnarray}
  Q(u) &\simeq& \frac{2 G m}{c^2 L^2 s} - \frac{2 G m}{c^2 L^2}
+ \frac{6 G m}{c^2}u^2 - \frac{8 G^2 m^2}{c^4}u^3
    \nonumber   \\
  &&+ \left(\frac{8 G^2 m^2}{c^4 L^2 s^2}
  -\frac{12 G^2 m^2}{c^4 L^2 s}+\frac{4 G^2 m^2}{c^4 L^2}\right)u\,,
\end{eqnarray}
where the $u^2$ corresponds to classical general relativity and
all other terms are contribution from the brane. Despite this
complicated form of $Q(u)$ one can solve analytically the
differential equation
\begin{equation}
  \frac{d^2 u}{d\phi^2} + u = Q\left(\frac{\cos\phi}{R}\right),
\end{equation}
which results in a rather lengthy expression. However, by
evaluating the solution $u(\phi = \pi/2 + \varepsilon) = 0$ and by
assuming that $\varepsilon$ is small, and performing a Taylor
expansion, the deflection angle becomes
\begin{equation}
  \delta = \frac{4GM}{c^2 R}\left(2-(1/s-1)\frac{R^2}{L^2}\right),
\end{equation}
where the factor up front corresponds to the case of general
relativity. Long baseline radio interferometry constrains the
factor in the bracket to be $\leq 1.0017$ and therefore $s$ would
have to be a very small real number to yield compatibility with
Solar System tests. This parameter choice would also contradict
the perihelion precession results discussed above.

Thus, there is no need to further discuss the radar echo delay for
this solution as it is already incompatible with Solar System
constraints. Lastly, we would like to note that these conclusions
also follow by considering the PPN parameters for the BMD
solution, which are given by $\gamma =2$ and $\beta= 2 + 4/s^2 -
2/s$. Clearly, $\gamma=2$ violates Solar System constrain while
the minimum value for $\beta$ is given by $\beta_{\rm min}=7/4$
which is attained when $s=4$, again incompatible with Solar System
tests.

\section{Discussions and final remarks}\label{SecV}

The study of the classical tests of general relativity provides a
very powerful method for constraining the allowed parameter space
of brane world solutions, and to provide a deeper insight into
the physical nature and properties of the corresponding spacetime
metrics. Therefore, this opens the possibility of testing
braneworld gravity by using astronomical and astrophysical
observations at the Solar System scale. In the present paper, we
have developed a general formalism that facilitates the analysis
of any given metric and provides the basic theoretical tools
necessary for the in depth comparison of the predictions of the
brane world models with the observational/experimental
results. In this context, the classical tests of general
relativity (perihelion precession, deflection of light, and the
radar echo delay) were considered for specific static and
spherically symmetric vacuum solutions in brane world models.

First, we analyzed a solution of the vacuum field equations,
obtained by Dadhich, Maartens, Papadopoulos and Rezania (DMPR)
\cite {Da00}, which represents the simplest generalization of the
Schwarzschild solution of general relativity. The tightest limit
we get on the parameter $Q$ came from the perihelion precession of
Mercury, and gives $\left| Q\right| \leq (1.32 \pm 1.63) \times
10^{30}$ MeV$^{-2}$, with other bounds about three orders of
magnitude larger. These results represent a significant
improvement of the results obtained in \cite{Boehmer:2008zh},
where a method based on the first order approximation of the
Hamilton-Jacobi result was used.

Second, we considered two families of analytic solutions,
parameterized by the ADM mass and the PPN parameters $\beta$ and
$\gamma$, found by Casadio, Fabbri and Mazzacurati (CFM)
\cite{cfm02}. These solutions can be physically seen as ``dark
pressure-dominated'' solutions. Contrary to what happens in the
DMPR case, here the most stringent bound came from the radar echo
delay data of the Cassini experiment, bounding the relevant
parameter $\eta$ to be $|\eta| \leq 0.000021 \pm 0.000024$.

Thus, these first two models analyzed are not directly ruled out by Solar System tests. To get a value that is more ``physically'' meaningful, we can use the matching conditions between the interior and the exterior of
a uniform density star in brane world, as derived in \cite{GeMa01}
to obtain the brane tension in terms of the relevant parameter of the
different solutions. We get:
\bea
\lambda_b &\geq& \frac{9GM_\odot^2}{4 \pi |Q| R_\odot^2} \nn \\
\lambda_b &\geq& \frac{27M_\odot c^2\left(1- \frac{3GM_\odot}{c^2R_\odot} \right)}
{8 \pi R_\odot^3 |\eta|} \nn \\
\lambda_b &\geq& \frac{9M_\odot c^2\left( \eta + \sqrt{1 -
\frac{2GM_\odot}{c^2R_\odot}(1+\eta)} \right)}{8 \pi R_\odot^3
|eta|(1+eta)} \label{tension} \eea for the DMPR, the CFM1 and the
CFM2 solutions respectively. Interestingly enough, the most
stringent bound comes from the DMPR solution (we are only
interested in orders of magnitude here, so we will omit actual
numerical values and errors): $\lambda_b \geq 10^7$ ${\rm GeV^4}$,
while the CFM solutions gives $\lambda_b \geq 10^4$ ${\rm GeV^4}$.
We do not have a definitive explanation for this, but we believe
it is related to the fact that in the CFM solutions the dark
radiation is basically negligible. Incidentally, the two different
bounds obtained in the different CFM solutions lead to exactly the
same bound on the brane tension, thus indicating that, although
causally different, the two solutions behave in the same way for
what concerns physical phenomena far away the central source.

In third place, we considered several classes of brane world black
hole solutions obtained by Bronnikov, Melnikov and Dehnen
\cite{BMD03} (BMD). The observational Solar System test place
stringent constraints on the parameter space of the model, and
these restrictions strongly violate Solar System constrains, so we
can safely rule out these solutions as viable solutions.

Despite the fact of having a whole plethora of brane world vacuum
solutions that pass the Solar System observational test a few
remarks are in order, namely, one may basically consider two
strategies of obtaining solutions on the brane. First, the bulk
spacetime may be given, by solving the full 5-dimensional
equations, and the geometry of the embedded brane is then deduced.
Second, due to the complexity of the 5-dimensional equations, one
may follow the strategy outlined in this paper, by considering the
intrinsic geometry on the brane, which encompasses the imprint
from the bulk, and consequently evolve the metric off the brane.
In principle, the second procedure may provide a well-determined
set of equations, with the brane setting the boundary data.
However, determining the bulk geometry proves to be an extremely
difficult endeavor.

\section*{Acknowledgments}

We would like to thank Maurizio Gasperini for helpful comments on
the manuscript. The work of TH is supported by the RGC grant No.
HKU 701808P of the government of the Hong Kong SAR. FSNL
acknowledges partial financial support of the Funda\c{c}\~{a}o
para a Ci\^{e}ncia e Tecnologia through the grants
PTDC/FIS/102742/2008 and CERN/FP/109381/2009.

\label{lastpage}

\end{document}